\documentclass[twocolumn,showpacs,amsmath,amssymb]{revtex4}

\usepackage{graphicx}
\usepackage{dcolumn}
\usepackage{bm}
\newcommand{\etal}{\textit{et al. }}
\newcommand{\half}{\textstyle \frac{1}{2}}
\begin{document}

\title{Electron transport through single Mn$_{12}$ molecular magnets}

\author{H. B. Heersche}
\email{hubert@qt.tn.tudelft.nl}
\author{Z. de Groot}
\author{J. A. Folk}
\thanks{{\it Present Address}: Department of Physics, University of British Columbia, Vancouver, Canada}
\author{H. S. J. van der Zant}

\affiliation{ Kavli Institute of Nanoscience, Delft University of Technology, Lorentzweg 1, 2628 CJ
Delft, The Netherlands}
\author{C. Romeike}
\author{M. R. Wegewijs}
\affiliation{
  Institut f\"ur Theoretische Physik A, RWTH Aachen, 52056 Aachen,  Germany
}
\author{L. Zobbi$^1$}
\author{D. Barreca$^2$}
\author{E. Tondello$^2$}
\author{A. Cornia$^1$}
\affiliation{$^1$Department of Chemistry, University of Modena and Reggio Emilia and INSTM, via G.
Campi 183, I-41100 Modena, Italy} \affiliation{$^2$ISTM-CNR, Department of Chemistry,  University
of Padova and INSTM,  Via Marzolo 1, I-35131 Padova, Italy}
\date{\today}

\begin{abstract}
We report transport measurements through a single-molecule magnet, the Mn$_{12}$ derivative
[Mn$_{12}$O$_{12}$(O$_2$C-C$_6$H$_4$-SAc)$_{16}$(H$_2$O)$_4$], in a single-molecule transistor
geometry. Thiol groups connect the molecule to gold electrodes that are fabricated by
electromigration. Striking observations are regions of complete current suppression and excitations
of negative differential conductance on the energy scale of the anisotropy barrier of the molecule.
Transport calculations, taking into account the high-spin ground state and magnetic excitations of
the molecule, reveal a blocking mechanism of the current involving non-degenerate spin multiplets.

\end{abstract}


\maketitle During the last few years it has been demonstrated that metallic contacts can
be attached to individual molecules allowing electron-transport measurements to probe
their intrinsic properties. Coulomb blockade and the Kondo effect, typical signatures of
a quantum dot, have been observed at low temperature in a single-molecule transistor
geometry~\cite{Park02,Liang02,Kubatkin03}. In the Coulomb blockade regime, vibrational
modes of the molecule appear as distinct features in the current-voltage ($I-V$)
characteristics~\cite{park_h00,Yu04vib,Pasupathy04}. Conformational, optical or magnetic
properties of molecules may also affect electron transport. With respect to the latter,
an interesting class of molecules is formed by the single-molecule magnets (SMMs). These
molecules show magnetic hysteresis due to their large spin and high anisotropy barrier,
which hampers magnetization reversal. The prototypal SMM, Mn$_{12}$-acetate, has total
spin $S=10$ and an anisotropy barrier of about 6~meV~\cite{Sessoli_93,Note5}.
 Spin excitations play an important role in the magnetization
reversal process through quantum tunneling of the magnetization (QTM)~\cite{Gatteschi_03}. Although
the electronic and magnetic properties of SMMs have been studied intensively on bulk samples and
thin films~\cite{Friedman_96,solut02, bisc05,nanol05,dunbar04}, the effect of the high-spin ground
state on electron transport in isolated molecules remains an unexplored topic~\cite{Kim04}.

\begin{figure}[b]
 \includegraphics[width=0.45\textwidth]{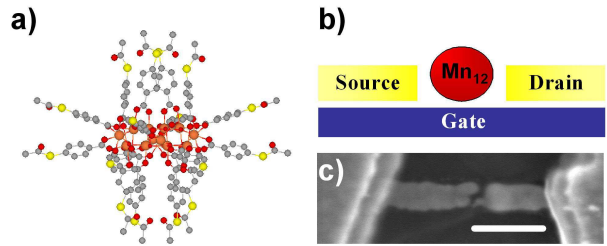}
  \caption{\label{fig:setup}
   (a) Side view of a Mn$_{12}$ molecule with tailor made ligands containing acetyl-protected thiol end groups (R=C$_6$H$_4$). Atoms are color
   labelled: Manganese (orange), oxygen (dark red), carbon (gray), sulfur
   (yellow). The diameter of the molecule is about 3~nm. (b) Schematic drawing of the Mn$_{12}$ molecule (red circle)
   trapped between electrodes. A gate can be used to change the
   electrostatic potential on the molecule enabling energy
   spectroscopy. (c) Scanning electron microscopy image of the electrodes. The gap is
   not resolvable. Scale-bar corresponds to 200~nm.}
\end{figure}

In this Letter we discuss transport through individual SMMs that are weakly coupled to
gold electrodes (see Fig.~\ref{fig:setup}). Experimental data show low-energy excited
states with a strong negative differential conductance (NDC) and regions of complete
current suppression (CCS). For comparison, we have modelled tunneling through a single
Mn$_{12}$ molecule taking into account its magnetic properties. Sequential tunnel
processes can result in spin-blockade of the current, providing a possible explanation
for the observed NDC and CCS. This effect is different from conventional spin
blockade~\cite{Weinmann95,Hutel_epl03} where there is no spin anisotropy.

We use the single-molecule magnets [Mn$_{12}$O$_{12}$(O$_2$C-R-SAc)$_{16}$(H$_2$O)$_4$]
(Mn$_{12}$ in the rest of the text), where R=$\{$C$_6$H$_4$, C$_{15}$H$_{30}$$\}$. Both
are tailor-made Mn$_{12}$-acetate derivatives~\cite{Note5}. These molecules feature thiol
groups in the outer ligand shell and consequently exhibit a strong affinity for gold
surfaces \cite{nanol05,angew03}. Beside ensuring robust tethering of the clusters to the
surface, the ligands are also believed to serve as tunnel barriers, so that the molecule
is only weakly coupled electronically to the gold. We assume that its magnetic properties
are conserved in the vicinity of gold electrodes. The diameter of the molecule (core plus
ligands) is about 3~nm for R=C$_6$H$_4$ and 5~nm for R= C$_{15}$H$_{30}$. In this paper
we focus on the transport features that are measured on the R=C$_6$H$_4$ derivative,
which is depicted in Fig.~\ref{fig:setup}(a).

\begin{figure}[b]
  \includegraphics[width=0.45\textwidth]{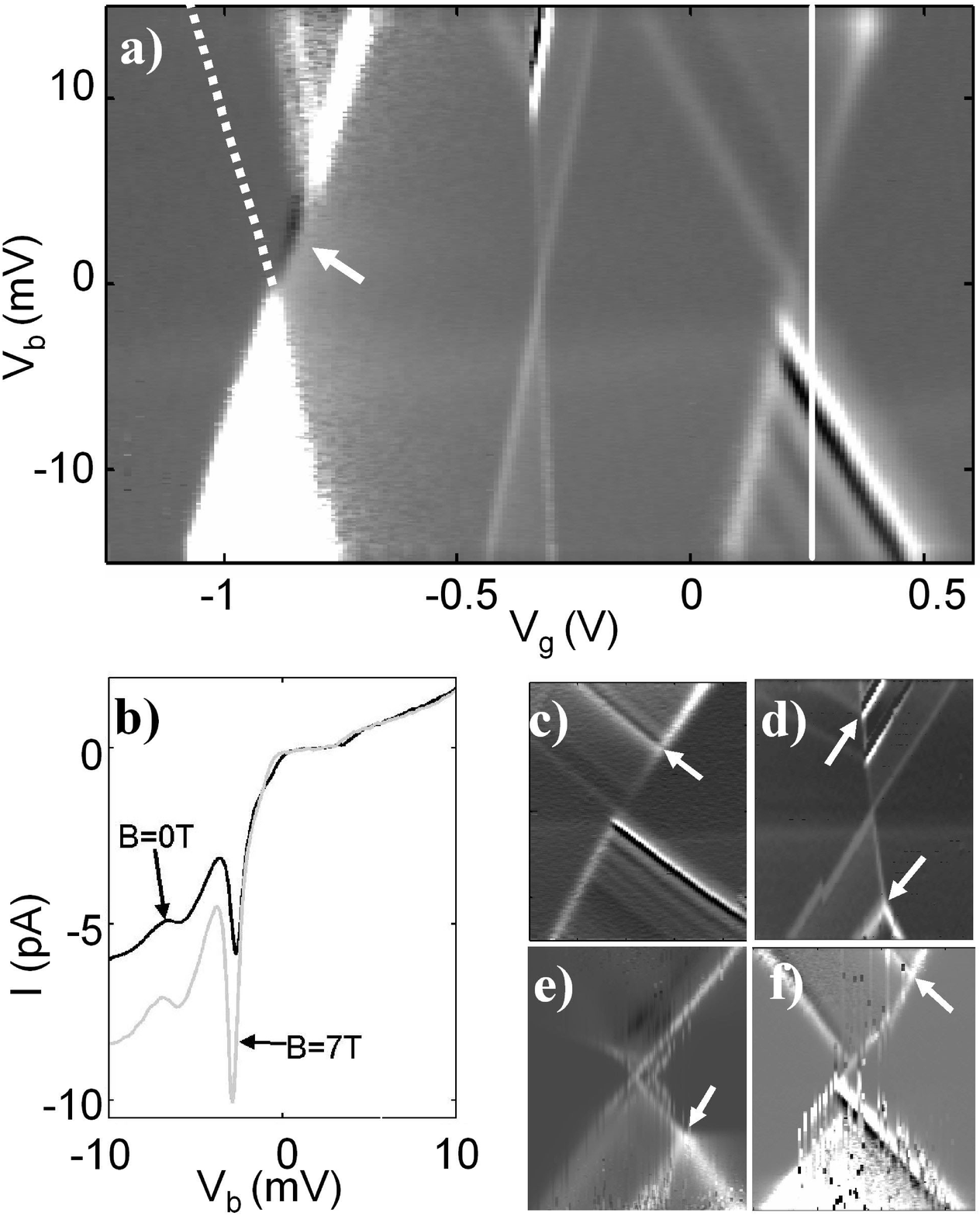}
  \caption{\label{fig:nice_fig1}
    (a)~Differential conductance (gray-scale) as a function of gate voltage
    ($V_g$) and bias voltage ($V_b$) ($T=3~$K, R=C$_6$H$_4$). A region of complete current suppression (left
    degeneracy point, arrow) and low-energy excitations with negative differential conductance (right degeneracy
    point) are observed. The dashed line near the left degeneracy point
    indicates the suppressed diamond edge. (gray-scale from -0.8~nS [black] to 1.4~nS [white]).
    (b)~$I-V_b$ at the gate voltage indicated in (a) with a line. NDC is
    clearly visible as a decrease in $|I|$ upon increasing $|V_b|$.
    Upon applying a magnetic field, current is increased for negative
    bias.(c-f) Same as (a) for 4 transport regions in which the 14~meV excitation is indicated with arrows. Bias voltage ranges are:
        $V_b=\pm$30~mV, $V_b=\pm20$~mV, $V_b=\pm25$~mV, $V_b=\pm15$~mV, respectively. (c,d) are the
        right and center charge transport regions of the sample in (a), respectively. (e,f) are
        measured in devices with the R=C$_{15}$H$_{30}$ ligands.}
\end{figure}

Electromigration~\cite{Park_h_apl1999} produced the nanometer-scale gaps in which the molecules
were trapped. We fabricated thin ($\sim 10$~nm) gold wires
 (width 100~nm, length 500~nm)
 on top of Al/Al$_2$O$_3$ gate electrodes using e-beam lithography. The wires were
contacted by thick (100~nm) gold leads. The samples were cleaned in acetone and iso-propanol,
descumed with an oxygen plasma, then soaked in a 0.1~mM Mn$_{12}$-solution containing a catalytic
amount of aqueous ammonia (to promote deprotection of the -SAc groups, see Ref.~\cite{angew03}) for
at least 1 hour. After taking a sample out of the solution it was dried in a nitrogen flux and
mounted in a He4 system with a $1$~K pot. The bridges were electromigrated in vacuum at room
temperature by ramping a voltage across the wire while monitoring the current using a series
resistor of 10~$\Omega$. The junctions broke at about 1~V.

After breaking, the samples were cooled down to $4$~K and the junction conductances were
measured as a function of gate voltage. Although about 10~$\%$ of 200 junctions showed
Coulomb blockade related features, only four samples were stable enough for detailed
measurements ~\cite{Note3}. In Fig.~\ref{fig:nice_fig1}(a) we plot the differential
conductance $G $ as a function of gate ($V_g$) and bias voltage ($V_b$) for one such a
device ($T=3$~K, R=C$_{6}$H$_{4}$). The lines separating the conducting regions from the
diamond-shaped Coulomb blockade regions have different slopes for the three different
charge transport regions. Within orthodox Coulomb blockade theory this implies that the
transport regions belong to different quantum dots, since the capacitance to the
environment is assumed constant for each dot. However, for molecular quantum dots it is
not possible to rule out that these three regions come from three different charge states
of the same molecule. Kubatkin \etal found that the charge distribution -and therefore
also the capacitance- of a (large) molecule depends on its charge
state~\cite{Kubatkin03}. In either scenario, however, the current in the non-overlapping
transport regions is determined by a single molecule.

\begin{figure}[t]
  \includegraphics[width=0.45\textwidth]{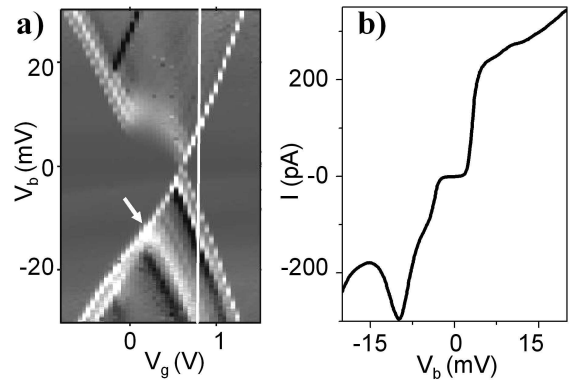}
  \caption{\label{fig:nice_fig2}
 (a)~Differential conductance (grayscale) as a function of gate voltage
    ($V_g$) and bias voltage ($V_b$) for another device ($T=1.5~$K, R=C$_6$H$_4$ ) in which the 2~meV and 3~meV excitations are
    also observed (negative bias). The 14~meV excitation is indicated by a white arrow. The origin of
    the step in the diamond edge at positive bias is unclear, but most likely not related to the
    magnetic properties of the molecule. (gray-scale from -3~nS [black] to 10~nS [white])
    (b)~$I-V_b$ along the line in (a).
  }
\end{figure}

Lines in transport regions running parallel to the diamond edges correspond to the onset
of transport through excited states of the molecule.  We have observed an excitation at
$14\pm1~$meV in all 6 of the stable transport regions that were observed in the four
samples. This excitation is indicated with arrows in Fig.~\ref{fig:nice_fig1}(c-f) and
Fig.~\ref{fig:nice_fig2}(a) and appears to be a fingerprint of the molecule. Its origin
can be vibrational since the Raman-spectrum is known to exhibit strong peaks beyond this
energy, which are
 associated with vibrations of the magnetic core of the molecule~\cite{Pederson02}.
A fingerprint is needed since small metal grains that are formed during electromigration
can mimic the transport properties of
single-molecules~\cite{Sordan_apl2005,Houck_nl05,Heersche_cond2005}. Moreover, we have
studied the electromigration process extensively and found that the gap size can be tuned
by the total (on- and off- chip) series resistance~\cite{zant_far06}. A relatively large
series resistance ($200~\Omega$) has been used to create a gap larger than $\sim 1~$ nm.
Using this technique, we did not measure any conductance up to 1~V in 50 control samples
without molecules deposited.

The focus of this paper is on transport features at low-energy ($\lesssim 5$~meV): a
region of complete current suppression (CCS) and a strong negative differential
conductance (NDC) excitation line in the stability diagrams. Both are visible in
Fig.~\ref{fig:nice_fig1}(a). At the left degeneracy point in this figure the current is
fully suppressed at positive bias voltage above the left diamond edge (dashed line).
Transport is restored beyond an excitation that lies at 5~meV. Remarkably, the right
diamond edge {\em does} continue all the way down to zero bias, defining a narrow strip
($\sim 1$~mV wide) where transport is possible.
 In the right conductive
regime in Fig.~\ref{fig:nice_fig1}(a), two excitations at an energy of 2~meV and 3~meV are the most
pronounced features. The 2~meV excitation is visible as a bright line with positive differential
conductance (PDC); the 3~meV excitation as a black line (NDC).
The strength of the  NDC is clearly visible in the ${I-V_b}$ plot in
Fig.~\ref{fig:nice_fig1}(b).

PDC and NDC excitations at 2~meV and 3~meV were also observed in a different device,  see
Fig.~\ref{fig:nice_fig2}.   Although the details of CCS and NDC varied from sample to
sample, these two features were consistent signatures of measurements in Mn$_{12}$. In
all cases, the 14~meV excitation is visible, suggesting that this feature originates in
the core of the molecule (which would not be so strongly affected by lead geometry). We
emphasize that we never observed low-energy features in bare gold samples or in samples
with other molecules deposited despite measuring over one thousand junctions in total.

\begin{figure}[tb]
   \includegraphics[width=0.5\textwidth]{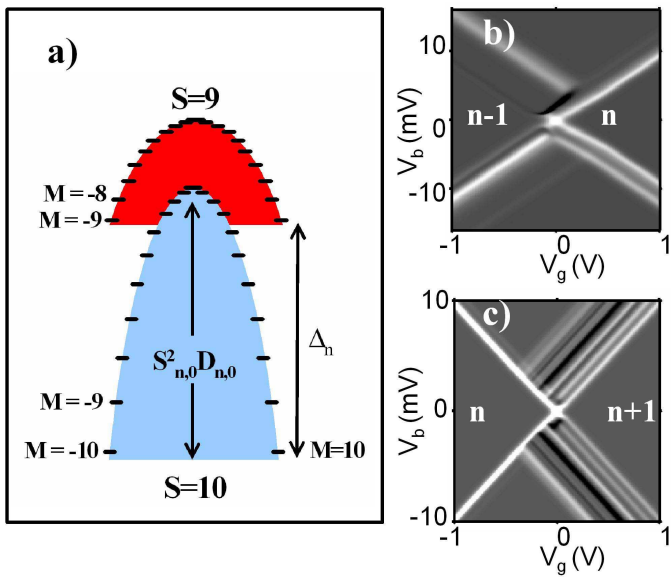}
 \caption{\label{fig:theory}
Model calculations. (a) Energy diagram of the $S_z$ states for the electronic ground
state ($\alpha =0$, blue) and first excited state ($\alpha =1$, red) of the neutral
molecule ($N=n$).  (b,c) Differential conductance (grayscale from {-0.8} [black] to 1
[white] a.u.) modeling CCS (b) and NDC (c). Parameters are listed in Ref.~\cite{Note4} }
\end{figure}

The observations of CSS and NDC lines at low energy that are characteristic of our
Mn$_{12}$ measurements do not follow in a straightforward way from conventional Coulomb
blockade theory.  This discrepancy should not be surprising, however, because Coulomb
blockade transport processes in a high-spin quantum dot (or molecule) have not yet been
been worked out in the literature.  For qualitative comparison to the experimental data,
we have developed a standard sequential tunneling model that incorparates the
spin-Hamiltonian description of the high-spin ground state of Mn$_{12}$ and its ladder of
spin excited states (see Fig.~\ref{fig:theory}). Both the total spin of the molecule $S$
and its projection $M$ on the intrinsic anisotropy axis of the molecule (the $z$-axis)
are taken into account.

This model provides one explanation for the observed NDC excitation and CCS.  Briefly:
the spin-selection rules $|\Delta S|, |\Delta M|= \half$ apply to adding or subtracting
an electron. Spin states of the molecule which differ by more than $\half$ from the
ground states are accessible via subsequent tunnel processes, but only if each step in
the sequence is energetically allowed. A sequence of tunnel processes can result in a
non-equilibrium population of certain excited states that can only be depopulated slowly
by a violation of the spin-selection rules induced by QTM. Transport is then hindered or
blocked at sufficiently low temperatures leading to NDC or CCS.

Our calculations (see Fig.~\ref{fig:theory}) take as a starting point the basic
spin-Hamiltonian for a SMM in charge state $N$ ( $N=n$ neutral, $N=n-1$ oxidized, $N=n+1$
reduced) and two $S$-states $\alpha=0,1$ for each charge state (energy splitting
$\Delta_N$) and total spin $S_{N\alpha}$:
\begin{equation}
  \label{hamil}
  H_{N \alpha}=-D_{N\alpha} S_{z}^2 + B_2 \left(S_x^2 -S_y^2 \right),
\end{equation}
where $D_{N \alpha}>0$ is an anisotropy constant and $B_2$ the
lowest order QTM perturbation due to deviations from perfect axial
symmetry. For $B_2=0$ this Hamiltonian gives rise to a ladder
consisting of the $2S_{N\alpha}+1$ different $M$ states, see
Fig.~\ref{fig:theory}(a).
 The states
$M=\pm S_{N 0}$ are degenerate ground states of the molecule and are
separated by the magnetic anisotropy barrier (MAB). Transition rates
between spin states upon adding or subtracting an electron are
determined by Clebsch-Gordan coefficients and spin selection rules.
We consider weak QTM (small $B_2$) where this picture is still
correct up to weak violations of the spin-selection rules which are
taken into account. Electronic- and spin-excitation relaxation
 rates are assumed to be much smaller than the tunnel
rates~\cite{wernsdorfer_epl00}.

Measurements show that the MAB  of the neutral Mn$_{12}$ molecule $(D_{n0}S^2_{n0})$
equals 5.6~meV ~\cite{Note5}. Recently, it has been demonstrated that the first excited
spin state in a neutral Mn$_{12}$ derivative has $S_{n,1}=9$ and lies about
$\Delta_n$=4~meV above the $S_{n,0}=10$ ground state~\cite{petukhov_prb04}, as
theoretically predicted by Park et al.~\cite{Park04add}. Little is known about
positively- or negatively- charged Mn$_{12}$ clusters~\cite{Park04add}, except that
one-electron reduced species have a $S_{n+1,0}=9\half$ ground state and a lower
MAB~\cite{Chakov_ioc2005}. We therefore have performed calculations for various values of
the remaining parameters and found that NDC and CCS are generic features whenever (i) the
MAB is charge state dependent (ii) the spin multiplets within a charge state overlap in
energy ($D_{N0}S_{N0}^2>\Delta_N$).

An example of CCS obtained from calculations is indicated in Fig.~\ref{fig:theory}(b) for
the $n-1 \leftrightarrow n$ transition and parameters listed in Ref.~\cite{Note4}. In
contrast to conventional spin-blockade, the current is not blocked in the narrow region
where only the ground states of the molecule are accessible. This is also observed
experimentally. Excitations with NDC are obtained for the $n \leftrightarrow n+1$
transition (see Fig~\ref{fig:theory}(c)). The mechanism is similar as for CCS, but
requires a lower MAB of the charged ground state spin multiplet~\cite{Note4}. It is
important to note that other parameter choices also give rise to CSS and NDC, so that a
quantitative comparison cannot be made at the moment.

Other explanations for NDC excitations and CCS have also been considered. Both intrinsic
and extrinsic (i.e. properties of the leads) properties could result in the observation
of NDC. The observation of an NDC excitation in different samples implies that it results
from an intrinsic property of the molecule, such as vibrational excitations. In some
rather special situations they are also predicted to result in
NDC~\cite{Koch04,Wegewijs05,Nowack04,Zazunov05}. A CCS feature of the type observed here,
however, has not been reported. Our model naturally gives rise to CCS and NDC due to the
intrinsic properties of the magnetic molecule: charge induced distortion of the magnetic
parameters and low-lying magnetic and spin-excitations.

The low-energy features of the different transport regions in Fig.~\ref{fig:nice_fig1}(a)
are not identical, possibly because they correspond to different charge state
transitions. In addition, the (magnetic) parameters may be effected by the different
coupling to the electrodes. Since the lines in the transport regions correspond to
sequences of transitions rather than to a single excitation, the pattern is sensitive to
variations of the magnetic parameters (and other parameters like temperature and tunnel
coupling).

The effect of a magnetic field in our transport situation is much less obvious than in
magnetic measurements on ensembles of molecules. Firstly, the angle of the external field
with respect to the easy axis of the molecule is unknown and cannot be controlled in our
experimental setup. A transverse field leads to mixing of spin eigenstates, allowing
transitions which are inhibited due to spin selection rules in the absence of a field,
whereas a longitudinal field will only shift the energies of the $S_z$ states. However,
estimates for the anisotropic g factors in the charge states are required to extend our
basic model. Furthermore, the electron tunnel couplings can be modified by the field. The
increase of the current upon applying a magnetic field, Fig.~\ref{fig:nice_fig1}(b), may
be related to all of the above effects but a full analysis of the magnetic field effects
is beyond the scope of the present paper.

In conclusion, we have measured transport through single Mn$_{12}$ molecules that are
weakly coupled to gold electrodes. Current suppression and negative differential
conductance on the energy scale of the anisotropy barrier have been observed. These
features can be understood qualitatively with a model, that combines the spin properties
of the molecule with standard sequential tunneling theory. These results provide a new
direction in the study of single-molecule magnets and possibly lead to electronic control
of nano-magnets.

HBH thanks H. Park's research group (Harvard) for their help and CR and MRW thank H.
Schoeller and J. Kortus for discussions. This work was supported by FOM, NWO (HBH, ZdG,
HSJvdZ), Pappalardo Postdoctoral Fellowship (JAF), the EU-RTN program on spintronics
(MRW), the HP Program RTN-QUEMOLNA, NOE"MAGMANET", and Italian MIUR (LZ, AC).

\bibliography{Heersche_Mn12_resub3}

\begin{thebibliography}{34}
\expandafter\ifx\csname natexlab\endcsname\relax\def\natexlab#1{#1}\fi
\expandafter\ifx\csname bibnamefont\endcsname\relax
  \def\bibnamefont#1{#1}\fi
\expandafter\ifx\csname bibfnamefont\endcsname\relax
  \def\bibfnamefont#1{#1}\fi
\expandafter\ifx\csname citenamefont\endcsname\relax
  \def\citenamefont#1{#1}\fi
\expandafter\ifx\csname url\endcsname\relax
  \def\url#1{\texttt{#1}}\fi
\expandafter\ifx\csname urlprefix\endcsname\relax\def\urlprefix{URL }\fi
\providecommand{\bibinfo}[2]{#2}
\providecommand{\eprint}[2][]{\url{#2}}

\bibitem[{\citenamefont{Park et~al.}(2002)}]{Park02}
\bibinfo{author}{\bibfnamefont{J.}~\bibnamefont{Park}} \bibnamefont{et~al.},
  \bibinfo{journal}{Nature} \textbf{\bibinfo{volume}{417}},
  \bibinfo{pages}{722} (\bibinfo{year}{2002}).

\bibitem[{\citenamefont{Liang et~al.}(2002)}]{Liang02}
\bibinfo{author}{\bibfnamefont{W.}~\bibnamefont{Liang}} \bibnamefont{et~al.},
  \bibinfo{journal}{Nature} \textbf{\bibinfo{volume}{417}},
  \bibinfo{pages}{725} (\bibinfo{year}{2002}).

\bibitem[{\citenamefont{Kubatkin et~al.}(2003)}]{Kubatkin03}
\bibinfo{author}{\bibfnamefont{S.}~\bibnamefont{Kubatkin}}
  \bibnamefont{et~al.}, \bibinfo{journal}{Nature}
  \textbf{\bibinfo{volume}{425}}, \bibinfo{pages}{698} (\bibinfo{year}{2003}).

\bibitem[{\citenamefont{Park et~al.}(2000)}]{park_h00}
\bibinfo{author}{\bibfnamefont{H.}~\bibnamefont{Park}} \bibnamefont{et~al.},
  \bibinfo{journal}{Nature} \textbf{\bibinfo{volume}{407}}, \bibinfo{pages}{57}
  (\bibinfo{year}{2000}).

\bibitem[{\citenamefont{Yu et~al.}(2004)}]{Yu04vib}
\bibinfo{author}{\bibfnamefont{L.~H.} \bibnamefont{Yu}} \bibnamefont{et~al.},
  \bibinfo{journal}{Phys.\ Rev.\ Lett.} \textbf{\bibinfo{volume}{93}},
  \bibinfo{pages}{266802} (\bibinfo{year}{2004}).

\bibitem[{\citenamefont{Pasupathy et~al.}(2005)}]{Pasupathy04}
\bibinfo{author}{\bibfnamefont{A.~N.} \bibnamefont{Pasupathy}}
  \bibnamefont{et~al.}, \bibinfo{journal}{Nano\ Lett.}
  \textbf{\bibinfo{volume}{5}}, \bibinfo{pages}{203} (\bibinfo{year}{2005}).

\bibitem[{\citenamefont{Sessoli et~al.}(1993)\citenamefont{Sessoli, Gatteschi,
  Caneschi, and Novak}}]{Sessoli_93}
\bibinfo{author}{\bibfnamefont{R.}~\bibnamefont{Sessoli}},
  \bibinfo{author}{\bibfnamefont{D.}~\bibnamefont{Gatteschi}},
  \bibinfo{author}{\bibfnamefont{A.}~\bibnamefont{Caneschi}}, \bibnamefont{and}
  \bibinfo{author}{\bibfnamefont{M.~A.} \bibnamefont{Novak}},
  \bibinfo{journal}{Nature} \textbf{\bibinfo{volume}{365}},
  \bibinfo{pages}{141} (\bibinfo{year}{1993}).

\bibitem[{Not({\natexlab{a}})}]{Note5}
\bibinfo{note}{See EPAPS Document No. [number will be inserted by publisher]
  for more information on the synthesis and analysis of the Mn$_{12}$
  (R=C$_6$H$_4$) samples.}

\bibitem[{\citenamefont{Gatteschi and Sessoli}(2003)}]{Gatteschi_03}
\bibinfo{author}{\bibfnamefont{D.}~\bibnamefont{Gatteschi}} \bibnamefont{and}
  \bibinfo{author}{\bibfnamefont{R.}~\bibnamefont{Sessoli}},
  \bibinfo{journal}{Angew.\ Chem.\ Int.\ Ed.} \textbf{\bibinfo{volume}{42}},
  \bibinfo{pages}{268} (\bibinfo{year}{2003}).

\bibitem[{\citenamefont{Friedman et~al.}(1996)\citenamefont{Friedman, Sarachik,
  Tejada, and Ziolo}}]{Friedman_96}
\bibinfo{author}{\bibfnamefont{J.~R.} \bibnamefont{Friedman}},
  \bibinfo{author}{\bibfnamefont{M.~P.} \bibnamefont{Sarachik}},
  \bibinfo{author}{\bibfnamefont{J.}~\bibnamefont{Tejada}}, \bibnamefont{and}
  \bibinfo{author}{\bibfnamefont{R.}~\bibnamefont{Ziolo}},
  \bibinfo{journal}{Phys.\ Rev.\ Lett.} \textbf{\bibinfo{volume}{76}},
  \bibinfo{pages}{3830} (\bibinfo{year}{1996}).

\bibitem[{\citenamefont{McInnes et~al.}(2002)}]{solut02}
\bibinfo{author}{\bibfnamefont{E.~J.~L.} \bibnamefont{McInnes}}
  \bibnamefont{et~al.}, \bibinfo{journal}{J. Am. Chem. Soc.}
  \textbf{\bibinfo{volume}{124}}, \bibinfo{pages}{9219} (\bibinfo{year}{2002}).

\bibitem[{\citenamefont{Cavallini et~al.}(2005)}]{bisc05}
\bibinfo{author}{\bibfnamefont{M.}~\bibnamefont{Cavallini}}
  \bibnamefont{et~al.}, \bibinfo{journal}{Angew. Chem. Int. Ed.}
  \textbf{\bibinfo{volume}{44}}, \bibinfo{pages}{888} (\bibinfo{year}{2005}).

\bibitem[{\citenamefont{Mannini et~al.}(2005)}]{nanol05}
\bibinfo{author}{\bibfnamefont{M.}~\bibnamefont{Mannini}} \bibnamefont{et~al.},
  \bibinfo{journal}{Nano Lett.} \textbf{\bibinfo{volume}{5}},
  \bibinfo{pages}{1435} (\bibinfo{year}{2005}).

\bibitem[{\citenamefont{Kim et~al.}(2004)}]{dunbar04}
\bibinfo{author}{\bibfnamefont{K.}~\bibnamefont{Kim}} \bibnamefont{et~al.},
  \bibinfo{journal}{Appl. Phys. Lett.} \textbf{\bibinfo{volume}{85}},
  \bibinfo{pages}{3872} (\bibinfo{year}{2004}).

\bibitem[{\citenamefont{Kim and Kim}(2004)}]{Kim04}
\bibinfo{author}{\bibfnamefont{G.-H.} \bibnamefont{Kim}} \bibnamefont{and}
  \bibinfo{author}{\bibfnamefont{T.-S.} \bibnamefont{Kim}},
  \bibinfo{journal}{Phys.\ Rev.\ Lett.} \textbf{\bibinfo{volume}{92}},
  \bibinfo{pages}{137203} (\bibinfo{year}{2004}).

\bibitem[{\citenamefont{Weinmann et~al.}(1995)\citenamefont{Weinmann,
  H{\"a}usler, and Kramer}}]{Weinmann95}
\bibinfo{author}{\bibfnamefont{D.}~\bibnamefont{Weinmann}},
  \bibinfo{author}{\bibfnamefont{W.}~\bibnamefont{H{\"a}usler}},
  \bibnamefont{and} \bibinfo{author}{\bibfnamefont{B.}~\bibnamefont{Kramer}},
  \bibinfo{journal}{Phys.\ Rev.\ Lett.} \textbf{\bibinfo{volume}{74}},
  \bibinfo{pages}{984} (\bibinfo{year}{1995}).

\bibitem[{\citenamefont{H\"{u}ttel et~al.}(2003)}]{Hutel_epl03}
\bibinfo{author}{\bibnamefont{H\"{u}ttel}} \bibnamefont{et~al.},
  \bibinfo{journal}{Eur.\ Phys.\ Lett.} \textbf{\bibinfo{volume}{62}},
  \bibinfo{pages}{712} (\bibinfo{year}{2003}).

\bibitem[{\citenamefont{Cornia et~al.}(2003)}]{angew03}
\bibinfo{author}{\bibfnamefont{A.}~\bibnamefont{Cornia}} \bibnamefont{et~al.},
  \bibinfo{journal}{Angew. Chem. Int. Ed.} \textbf{\bibinfo{volume}{42}},
  \bibinfo{pages}{1645} (\bibinfo{year}{2003}).

\bibitem[{\citenamefont{Park et~al.}(1999)}]{Park_h_apl1999}
\bibinfo{author}{\bibfnamefont{H.}~\bibnamefont{Park}} \bibnamefont{et~al.},
  \bibinfo{journal}{Appl. Phys. Lett.} \textbf{\bibinfo{volume}{75}}
  (\bibinfo{year}{1999}).

\bibitem[{Not({\natexlab{b}})}]{Note3}
\bibinfo{note}{The yield was similar for both types of derivatives. Two of the
  stable samples where obtained with the R=C$_{15}$H$_{30}$ derivative.}

\bibitem[{\citenamefont{Pederson et~al.}(2002)\citenamefont{Pederson,
  Bernstein, and Kortus}}]{Pederson02}
\bibinfo{author}{\bibfnamefont{M.~R.} \bibnamefont{Pederson}},
  \bibinfo{author}{\bibfnamefont{N.}~\bibnamefont{Bernstein}},
  \bibnamefont{and} \bibinfo{author}{\bibfnamefont{J.}~\bibnamefont{Kortus}},
  \bibinfo{journal}{Phys.\ Rev.\ Lett.} \textbf{\bibinfo{volume}{89}},
  \bibinfo{pages}{097202} (\bibinfo{year}{2002}).

\bibitem[{\citenamefont{Sordan et~al.}(2005)\citenamefont{Sordan,
  Balasubramanian, Burghard, and Kern}}]{Sordan_apl2005}
\bibinfo{author}{\bibfnamefont{R.}~\bibnamefont{Sordan}},
  \bibinfo{author}{\bibfnamefont{K.}~\bibnamefont{Balasubramanian}},
  \bibinfo{author}{\bibfnamefont{M.}~\bibnamefont{Burghard}}, \bibnamefont{and}
  \bibinfo{author}{\bibfnamefont{K.}~\bibnamefont{Kern}},
  \bibinfo{journal}{Appl. Phys. Lett.} \textbf{\bibinfo{volume}{87}},
  \bibinfo{pages}{013106} (\bibinfo{year}{2005}).

\bibitem[{\citenamefont{Houck et~al.}(2005)}]{Houck_nl05}
\bibinfo{author}{\bibfnamefont{A.~A.} \bibnamefont{Houck}}
  \bibnamefont{et~al.}, \bibinfo{journal}{Nano\ Lett.}
  \textbf{\bibinfo{volume}{5}}, \bibinfo{pages}{1685} (\bibinfo{year}{2005}).

\bibitem[{\citenamefont{Heersche et~al.}(2006)}]{Heersche_cond2005}
\bibinfo{author}{\bibfnamefont{H.~B.} \bibnamefont{Heersche}}
  \bibnamefont{et~al.}, \bibinfo{journal}{Phys.\ Rev.\ Lett.}
  \textbf{\bibinfo{volume}{96}}, \bibinfo{pages}{017205}
  (\bibinfo{year}{2006}).

\bibitem[{\citenamefont{van~der Zant et~al.}(2006)}]{zant_far06}
\bibinfo{author}{\bibfnamefont{H.}~\bibnamefont{van~der Zant}}
  \bibnamefont{et~al.}, \bibinfo{journal}{Faraday discussions}
  \textbf{\bibinfo{volume}{131}}, \bibinfo{pages}{347} (\bibinfo{year}{2006}).

\bibitem[{\citenamefont{Wernsdorfer et~al.}(2000)}]{wernsdorfer_epl00}
\bibinfo{author}{\bibfnamefont{W.}~\bibnamefont{Wernsdorfer}}
  \bibnamefont{et~al.}, \bibinfo{journal}{Eur.\ Phys.\ Lett.}
  \textbf{\bibinfo{volume}{50}}, \bibinfo{pages}{552} (\bibinfo{year}{2000}).

\bibitem[{\citenamefont{Petukhov et~al.}(2004)}]{petukhov_prb04}
\bibinfo{author}{\bibfnamefont{K.}~\bibnamefont{Petukhov}}
  \bibnamefont{et~al.}, \bibinfo{journal}{Phys.\ Rev.\ B}
  \textbf{\bibinfo{volume}{70}}, \bibinfo{pages}{054426}
  (\bibinfo{year}{2004}).

\bibitem[{\citenamefont{Park and Pederson}(2004)}]{Park04add}
\bibinfo{author}{\bibfnamefont{K.}~\bibnamefont{Park}} \bibnamefont{and}
  \bibinfo{author}{\bibfnamefont{M.~R.} \bibnamefont{Pederson}},
  \bibinfo{journal}{Phys.\ Rev.\ B} \textbf{\bibinfo{volume}{70}},
  \bibinfo{pages}{054414} (\bibinfo{year}{2004}).

\bibitem[{\citenamefont{Chakov et~al.}(2005)}]{Chakov_ioc2005}
\bibinfo{author}{\bibfnamefont{N.}~\bibnamefont{Chakov}} \bibnamefont{et~al.},
  \bibinfo{journal}{Inorg. Chem.} \textbf{\bibinfo{volume}{44}},
  \bibinfo{pages}{5304} (\bibinfo{year}{2005}).

\bibitem[{Not({\natexlab{c}})}]{Note4}
\bibinfo{note}{Model parameters (energies in [meV]) $(S_{n,0},S_{n,1})=(10,
  9)$, $(D_{n,0},D_{n,1})=(0.056, 0.04)$ $\Delta_n=4.0$ and $B_2=10^{-4}$.
  Parameters for CCS: $(S_{n-1,0},S_{n-1,1})=(9\half, 9\half)$,
  $(D_{n-1,0},D_{n-1,1})=(0.02 , 0.05)$, $\Delta_{n-1}=1.3$,
  $\Gamma_L/\Gamma_R=2$, and $T= 2$~K. Parameters for NDC:
  $(S_{n+1,0},S_{n+1,1})=(9\half, 9\half)$, $(D_{n+1,0},D_{n+1,1})=(0.02,
  0.02)$, $\Delta_{n+1}=0.6$, $B_2=10^{-5}$, $\Gamma_L/\Gamma_R=10$, and $T=
  1$~K. The NDC {\em strength} can only be reproduced for a temperature lower
  than the experimental one in Fig.~\ref{fig:nice_fig1},~\ref{fig:nice_fig2}}.

\bibitem[{\citenamefont{Koch and von Oppen}(2005)}]{Koch04}
\bibinfo{author}{\bibfnamefont{J.}~\bibnamefont{Koch}} \bibnamefont{and}
  \bibinfo{author}{\bibfnamefont{F.}~\bibnamefont{von Oppen}},
  \bibinfo{journal}{Phys.\ Rev.\ Lett.} \textbf{\bibinfo{volume}{94}},
  \bibinfo{pages}{206804} (\bibinfo{year}{2005}).

\bibitem[{\citenamefont{Wegewijs and Nowack}(2005)}]{Wegewijs05}
\bibinfo{author}{\bibfnamefont{M.~R.} \bibnamefont{Wegewijs}} \bibnamefont{and}
  \bibinfo{author}{\bibfnamefont{K.~C.} \bibnamefont{Nowack}},
  \emph{\bibinfo{title}{Focus on NEMS}} (\bibinfo{publisher}{New J. Phys.},
  \bibinfo{year}{2005}), vol.~\bibinfo{volume}{7}, p. \bibinfo{pages}{239}.

\bibitem[{\citenamefont{Nowack and Wegewijs}(2004)}]{Nowack04}
\bibinfo{author}{\bibfnamefont{K.~C.} \bibnamefont{Nowack}} \bibnamefont{and}
  \bibinfo{author}{\bibfnamefont{M.~R.} \bibnamefont{Wegewijs}}
  (\bibinfo{year}{2004}), \bibinfo{note}{cond-mat/0506552}.

\bibitem[{\citenamefont{Zazunov et~al.}()\citenamefont{Zazunov, Feinberg, and
  Martin}}]{Zazunov05}
\bibinfo{author}{\bibfnamefont{A.}~\bibnamefont{Zazunov}},
  \bibinfo{author}{\bibfnamefont{D.}~\bibnamefont{Feinberg}}, \bibnamefont{and}
  \bibinfo{author}{\bibfnamefont{T.}~\bibnamefont{Martin}},
  \bibinfo{note}{cond-mat/0510066}.

\end{thebibliography}
\end{document}